\documentclass[11pt,a4paper,oneside]{article}
\usepackage[T1]{fontenc}
\usepackage[ansinew]{inputenc}
\usepackage[english]{babel}
\usepackage{amsfonts}
\usepackage{amsmath}
\usepackage{bm}
\usepackage{array}
\usepackage{amsthm}
\usepackage{amssymb}
\usepackage{graphicx}
\usepackage{braket}
\usepackage{verbatim}
\usepackage[table]{xcolor}
\usepackage{caption}
\usepackage{cite}
\usepackage{textcomp}
\usepackage{url}
\usepackage{hyperref}
\usepackage{mathtools}

\hypersetup{
    colorlinks=true,
    linkcolor=black,
    citecolor=black
    }
\usepackage{multicol}
\newcommand{\be}{\begin{equation}}
\newcommand{\ee}{\end{equation}}
\newcommand{\beqn}{\begin{eqnarray}}
\newcommand{\eeqn}{\end{eqnarray}}
\definecolor{pinegreen}{rgb}{0.0, 0.47, 0.44}

\renewcommand{\d}{\mbox{${\rm d}$}}

\newtheorem{theorem}{Theorem}
\newtheorem{corollary}{Corollary}

\theoremstyle{definition}

\theoremstyle{remark}

\newcommand{\K}{\mathcal{K}}
\newcommand{\T}{\mathcal{T}}
\title{\bf Thermal aspects of the anomalous\\ $\omega \to \infty$ limit of Brans-Dicke gravity}
\author{L.~Gallerani$^{a}$\thanks{E-mail: luca.gallerani6@unibo.it},
$\ $
A.~Giusti$^{bc}$\thanks{Email: andrea.giusti9@unibo.it},
$\ $
A.~Mentrelli$^{ab}$\thanks{E-mail: andrea.mentrelli@unibo.it},
$\ $
V.~Faraoni$^{d}$\thanks{E-mail: vfaraoni@ubishops.ca}
$\ $
\\
\\
$^a${\em Department of Mathematics \& ${\cal AM}^2$, University of Bologna, 40123 Bologna, Italy}
\\
\\
$^b${\em I.N.F.N., Sezione di Bologna, I.S.~FLAG, 40127 Bologna, Italy}
\\
\\
$^c${\em DIFA \& ${\cal AM}^2$, University of Bologna, 40126 Bologna, Italy}
\\
\\
$^d${\em Department of Physics \& Astronomy, Bishop's University}\\
{\em 2600 College Street, Sherbrooke, Qu{\'e}bec, Canada J1M 1Z7}
}
\begin{document}
\maketitle
\begin{abstract}
\noindent Brans-Dicke gravity does not always reduce to General Relativity in the limit $\omega\to\infty$ for the coupling constant. This anomalous behavior is examined within the formalism of the first-order thermodynamics of scalar-tensor gravity. It is shown that this effect is linked to the non-vanishing nature of the chemical potential, in the Einstein frame formulation of the thermodynamic analogy, in the $\omega\to\infty$ limit.
\end{abstract}

\newpage

\section{Introduction and main results}
\label{sec:intro}
Brans-Dicke (BD) gravity \cite{Brans:1961sx} is a generalization of General Relativity (GR) that introduces a non-minimally coupled scalar field $\phi$ to the gravitational sector of the Einstein-Hilbert action. This scalar degree of freedom then plays the role of an effective gravitational coupling $G_{\rm eff} \simeq \phi^{-1}$ for the theory.

The BD action, in the Jordan conformal frame, reads 
\be
\label{SBD}
S_{\rm BD}=\frac{1}{16\pi}\int d^4 x \sqrt{-g}\left[\phi R 
-\frac{\omega}{\phi} \, \nabla_c\phi\nabla^c\phi 
-V(\phi)\right]+S^{(\mathrm{m})}
\ee
where $\phi$ is the BD scalar, $\omega$ denotes the BD coupling, $V(\phi)$ is the BD potential for $\phi$, and $S^{(m)}$ is the matter action. The field equations associated with this action then read
\beqn
R_{ab}-\frac{1}{2}g_{ab}R&\!\!=\!\!&\frac{8\pi}{\phi}T^{(\mathrm{m})}_{ab}+\frac{\omega}{\phi^2}\left(\nabla_a\phi\nabla_b\phi-\frac{1}{2}g_{ab}\nabla_c\phi\nabla^c\phi\right) +\frac{1}{\phi}(\nabla_a\nabla_b\phi-g_{ab}\Box\phi)-\frac{V}{2\phi}g_{ab} \, , \quad \label{JBD1} \\
\Box\phi &\!\!=\!\!& \frac{1}{2\omega+3}\left({8\pi T^{(m)}}+\phi \, V' - 2 V\right) \, , \label{JBD2}
\eeqn
where $R_{ab}$ and $R$ are the Ricci tensor and Ricci scalar respectively, and $\nabla_a$
denotes the covariant derivative associated with the Levi-Civita connection compatible with the metric $g_{ab}$, $\Box:= g^{ab}\nabla_a \nabla_b$, $V' := dV/d\phi$ and $T^{(\mathrm{m})}=g^{cd}T^{(\mathrm{m})}_{cd}$ is the trace of the matter energy-momentum tensor. Note that we adopt the notation of Ref.~\cite{Wald:1984rg}, while also setting the speed of light and the Newtonian constant of gravitation to unity.

It is clear from the above expressions that BD gravity reduces to GR when the scalar field is taken to be constant. A common belief was that the same GR limit for the theory could be achieved by formally taking $\omega \to \infty$, as popularized in some celebrated textbook such as \cite{Weinberg:1972kfs}. However, it was later shown \cite{Faraoni:1999yp,Banerjee:1996iy,Miyazaki:2000ij,Brando:2018kic,Faraoni:2019sxw} that this is not generally true and that the asymptotic behavior of $\phi$, as a function of $\omega$, plays a pivotal role in determining whether this statement holds. In particular, although it is generally true that if $\phi\sim\phi_{\infty}+\mathcal{O}(1/\omega)$, with $\phi_{\infty}$ constant, BD gravity will relax to GR as $\omega \to \infty$ (see \cite{Weinberg:1972kfs}), several counterexamples to the generality of this statement were found in terms of exact solutions for {\em electrovacuum} BD gravity featuring an asymptotic behavior such that
$$
\phi \sim \phi_{\infty}+\mathcal{O} \left( \frac{1}{\sqrt{|\omega|}}\right) \, , \quad \mbox{as} \,\, \omega \to \infty \, ,
$$
see e.g.~\cite{Faraoni:1999yp,Banerjee:1996iy,Miyazaki:2000ij,Brando:2018kic}.\footnote{Notably, this anomalous behavior is also present in the case of BD gravity coupled to matter fields \cite{Nguyen:2024jgy}.} This weakens the significance of the constraints of Solar System experiments for scalar-tensor theories \cite{Faraoni:2019sxw}, at least for those whose conclusions rely upon the strict connection between $\omega \to \infty$ and the GR limit for BD gravity.  

The {\em thermodynamics of scalar-tensor gravity} \cite{Faraoni:2021lfc,Faraoni:2021jri,Giusti:2021sku} (see also \cite{Giardino:2023ygc} for a short review) interprets the second piece in the right-hand side of Eq.~\eqref{JBD1} as the stress-energy tensor of an {\em effective fluid}, i.e.
\be
\label{Tphiab}
8\pi T^{(\phi)}_{ab}=\frac{\omega}{\phi^2}\left(\nabla_a\phi\nabla_b\phi-\frac{1}{2}g_{ab}\nabla_c\phi\nabla^c\phi\right)+\frac{1}{\phi}(\nabla_a\nabla_b\phi-g_{ab}\Box\phi)-\frac{V}{2\phi}g_{ab} \, ,
\ee
and, taking advantage of the {\em imperfect fluid decomposition} \cite{Deffayet:2010qz,Faraoni:2018nql} for $T^{(\phi)}_{ab}$, allows for the derivation of the constitutive laws of the effective fluid, dubbed {\em $\phi$-fluid}. More precisely, let $\nabla _a \phi$ be timelike, then we define the 4-velocity of the $\phi$-fluid as
\be
\label{gvelocity}
u^{a} :=\epsilon \, \frac{\nabla^a\phi}{\sqrt{2X}}
\ee
with $2X := -\nabla_c \phi \nabla^c\phi$, $u_a u^a=-1$, and with $\epsilon=+1$ if $\nabla _a \phi$ is future-oriented and $\epsilon=-1$ otherwise (this choice of $\epsilon$ guarantees that $u^a$ is always future-oriented, see \cite{Giusti:2022tgq}). The effective stress-energy tensor $T^{(\phi)}_{ab}$ admits the imperfect fluid decomposition
\be 
T^{(\phi)} _{ab}=\rho^{(\phi)} \, u_a u_b + q^{(\phi)} _{a} u_{b} + q^{(\phi)} _{b} u_{a} + P^{(\phi)} \, h_{ab} + \pi^{(\phi)} _{ab} \, ,
\ee
with $\rho^{(\phi)}$ the effective energy density, $q^{(\phi)} _{a}$ the effective heat-flux density, $P^{(\phi)}$ the effective isotropic pressure, $h_{ab} := g_{ab} + u_a u_b$, and $\pi^{(\phi)} _{ab}$ the effective (anisotropic) stress tensor. Explicit expressions for the aforementioned quantities can be found, for the case of BD gravity and its simplest generalization, in \cite{Faraoni:2021lfc,Faraoni:2021jri}. Relating the explicit expressions for the components of the above decomposition with the kinematic quantities associated to the $\phi$-fluid (i.e., the expansion scalar, the shear and rotation tensors) one finds the constitutive relations for the effective fluid. Surprisingly, one can draw an analogy between such constitutive laws and those of Eckart's irreversible thermodynamics of continuous media \cite{Eckart:1940te}. This procedure then naturally yields a formal notion of temperature $\T$ and thermal conductivity $\K$ for scalar-tensor theories of gravity \cite{Faraoni:2021lfc,Faraoni:2021jri,Giusti:2021sku}. For BD gravity, specifically, one has that
\be 
\label{KT}
\K \T=\frac{\sqrt{-\nabla_a\phi\nabla^a\phi}}{8\pi \phi} \, .
\ee
Different choices of $T^{(\phi)}_{ab}$ lead to slightly different definitions of temperature, nonetheless conclusions drawn from all these alternative formulations of the analogy are physically independent of this choice (see \cite{Gallerani:2024gdy}). 

Interestingly, in the thermodynamics of scalar-tensor gravity one can derive a generalized heat equation governing the evolution of $\K \T$, indeed for BD gravity one can easily find that \cite{Faraoni:2021jri}
\be 
\frac{d (\K \T)}{\d \tau} := u^a \nabla_a \left(\K \T \right) = 8\pi \left( {\cal KT}\right)^2 
-\Theta {\cal KT} +\frac{ \Box\phi}{8\pi \phi} \, . \label{evolution_general} 
\ee 
This equation then determines the stability of a thermal state, and how far it is from equilibrium. The generalized heat equation, combined with Eq.~\eqref{KT}, then allows us to conclude that GR (i.e., BD gravity with $\phi = {\rm const.}$) corresponds to a zero-temperature {\em equilibrium} state within this formalism, namely
\be
\K \T \Big|_{\rm GR} = 0 \, .
\ee
Notably, this framework also allows one to investigate the tendency of a given scalar-tensor model to approach GR over time \cite{Faraoni:2025alq}, thus determining a ``thermal origin'' for the famous  Damour-Nordvedt attractor mechanism to GR \cite{Damour:1992kf}. 

The results discussed so far hold for scalar-tensor theories formulated in the Jordan conformal frame. It was later shown in \cite{Faraoni:2022gry} that the thermal analogy can be extended also to the Einstein frame. However, since the scalar field becomes minimally coupled to the Einstein-Hilbert term in this frame, the Fourier law that allowed to determine the temperature in the Jordan frame get mapped into Fick's law in the Einstein frame. In other words, the analogy with Eckart's theory in the Jordan frame ($\T \neq 0$ and no particle diffusion) gets mapped into particle diffusion at zero temperature in the Einstein frame. The role of $\K \T$ in the Einstein frame is then taken on by the chemical potential, which reads
\be 
\label{eq:defChem}
\widetilde{\mu} = \sqrt{2 \, \widetilde{X}} \, ,
\ee
with $2\widetilde{X} := -\widetilde{\nabla}_c \widetilde{\phi} \, \widetilde{\nabla}^c \widetilde{\phi}$ denoting the Lorentzian norm of the scalar field expressed in terms of Einstein-frame quantities.

In this work we prove the following statements.
\begin{theorem}\label{Thm-1}
Consider {\em vacuum} Brans-Dicke gravity (i.e., the action in \eqref{SBD} with $S^{\rm (m)} = 0$) in the Jordan frame. Let $(M,g,\phi)$ be a smooth solution of the {\em vacuum} BD field equations
\beqn
R_{ab}-\frac{1}{2}g_{ab}R &\!\!=\!\!& \frac{\omega}{\phi^2}\left(\nabla_a\phi\nabla_b\phi-\frac{1}{2}g_{ab}\nabla_c\phi\nabla^c\phi\right) +\frac{1}{\phi}(\nabla_a\nabla_b\phi-g_{ab}\Box\phi)-\frac{V}{2\phi}g_{ab} \, , \quad \label{JBD1-v} \\
\Box\phi &\!\!=\!\!& \frac{1}{2\omega+3}\left(\phi \, V' - 2 V\right) \, , \label{JBD2-c}
\eeqn
with $(M,g)$ denoting the spacetime manifold and $\phi$ the BD scalar. Furthermore, assume that $(g,\phi)$ admit the expansions 
\beqn
\phi (x) &\!\!=\!\!& \phi_{\infty} + \frac{\Phi (x)}{\sqrt{|\omega|}} + \mathcal{O} \left( 
\frac{1}{|\omega|^{\frac{1}{2}+\alpha}} \right) \, , \label{A1} \\
g_{ab} (x) &\!\!=\!\!& g^{(\infty)} _{ab} (x) + \mathcal{O} \left( 
\frac{1}{|\omega|^\beta} \right) \, \label{A2} ,
\eeqn
as $|\omega| \to \infty$, allowing for term-wise differentiation up to second order in the spacetime variables, where $\Phi$ and $g^{(\infty)} _{ab}$ are respectively some smooth scalar field, with timelike future-oriented gradient, and a nondegenerate rank-2 symmetric tensor field with Lorentzian signature, both independent of $\omega$, $\phi_\infty$ is a real constant, and $\alpha,\beta > 0$. Then the {\em Einstein frame chemical potential} satisfies
\be 
\lim _{|\omega| \to  \infty} \tilde{\mu} \neq 0 \, .
\ee
\end{theorem}
\begin{corollary}
\label{Corollary}
In the assumptions of {\bf Theorem \ref{Thm-1}}, the GR limit of {\em vacuum} BD gravity and the $|\omega| \to \infty$ limit do not coincide. 
\end{corollary}

\section{Jordan frame {\em vs.} Einstein frame}
\label{sec:JFEF}
In this section we review the main properties of the map relating Jordan frame and Einstein frame formulations of scalar-tensor gravity. For further details we refer the reader to \cite{Faraoni:1999hp} and references therein.

Let $(g_{ab},\phi)$ denote the {\em Jordan frame variables} for BD theory, defined by the action in \eqref{SBD}. The map of the theory to the Einstein frame consists of two steps:
\begin{itemize}
    \item[(i)] A conformal transformation of the metric tensor:
    \be
    g_{ab}\longrightarrow \widetilde{g}_{ab}=\phi \, g_{ab} \, ;
    \ee
    \item[(ii)] a field redefinition:
    \be
    \label{phiE}
    \phi \longrightarrow\tilde{\phi}=\sqrt{\frac{|2\omega +3|}{16\pi}}\ln{\left(\frac{\phi}{\phi_0}\right)},
    \ee
    where $\phi_0$ is a constant with the same physical dimensions as $\phi$.  
\end{itemize}
The pair $(\widetilde{g}_{\mu \nu}, \widetilde{\phi})$ then denotes the Einstein frame variables for the theory.

Observing that
\be
\label{Covariant}
\nabla_{a}\phi=\phi_0\sqrt{\frac{16\pi}{|{2\omega+3}|}}\exp{\left(\sqrt{\frac{16\pi}{|2\omega+3|}}\widetilde{\phi}\right)}\widetilde{\nabla}_{a}\widetilde{\phi} \, ,
\ee
one can then reformulate the BD action~\eqref{SBD} in the Einstein frame as:
\be
S^{\rm (E)}_{\rm BD} = \int d^4 x \sqrt{-\tilde{g}}\left[\frac{\tilde{R}}{16\pi}-\frac{1}{2}\tilde{g}^{ab}\tilde{\nabla}_a\tilde{\phi}\tilde{\nabla}_b\tilde{\phi}-U(\tilde{\phi})+\frac{\mathcal{L}^{(\mathrm{m})}}{\phi^2(\tilde{\phi})}\right] \, , \qquad 
U(\tilde{\phi}):=\frac{V(\phi)}{16\pi \phi^2}\Bigr|_{\phi=\phi(\tilde{\phi})} \,\, .
\ee
From the expression of the Einstein-frame action for BD gravity one can immediately conclude that the Einstein-frame BD scalar field is both minimally coupled to the Einstein-Hilbert term and coupled non-minimally to the matter fields. Furthermore, the Einstein-frame field equations for BD gravity read:
\beqn
&& \tilde{R}_{ab}-\frac{1}{2}\tilde{g}_{ab}\tilde{R} =
8\pi \left(e^{-\sqrt{\frac{64\pi}{|2\omega+3|}}\tilde{\phi}}T^{(\mathrm{m})}_{ab}+\tilde{\nabla}_a\tilde{\phi}\tilde{\nabla}_b\tilde{\phi}-\frac{1}{2}\tilde{g}_{ab}\tilde{g}^{cd}\tilde{\nabla}_c\tilde{\phi}\tilde{\nabla}_d\tilde{\phi}-U(\tilde{\phi})\tilde{g}_{ab}\right) \, , \label{EBD1} \\
&&
\tilde{\Box} \tilde{\phi}-\frac{dU}{d\tilde{\phi}}+8\sqrt{\frac{\pi}{|2\omega+3|}}e^{-\sqrt{\frac{64\pi}{|2\omega+3|}}\tilde{\phi}}\mathcal{L}^{(\mathrm{m})}=0 \, . \label{EBD2} 
\eeqn
\section{Proof of Theorem~\ref{Thm-1}}
\label{sec:proof}
As discussed in \cite{Faraoni:2019sxw}, the origin of the anomalous limit of {\em vacuum} BD gravity can be traced back to the fact that the term
\be
A_{ab} := \frac{\omega}{\phi^2}\left(\nabla_a\phi \nabla_b\phi -\frac{1}{2}g_{ab}\nabla_c\phi\nabla^c\phi\right) \, ,
\ee
in the Jordan frame field equations does not vanish as $|\omega| \to \infty$ if 
$$ 
\phi \sim \phi_{\infty} + \mathcal{O} (1/\sqrt{|\omega|}) \, .$$ 
More precisely, given the above asymptotic behavior and the Jordan-Einstein frame map discussed in Sec.~\ref{sec:JFEF}, we find that
\beqn
A_{ab} \xrightarrow[]{|\omega| \to \infty}\!\!\!\! &&A^{(\infty)}_{ab} = 8\pi \, {\rm sign}(\omega)\left( \tilde{\nabla}_a \tilde{\phi}\tilde{\nabla}_b \tilde{\phi}-\frac{1}{2}\tilde{g}_{ab} \, \tilde{g}^{cd} \tilde{\nabla}_c \tilde{\phi}\tilde{\nabla}_d \tilde{\phi}\right) \, , \\
g_{ab} = \phi^{-1} \, \tilde{g}_{ab} \xrightarrow[]{|\omega| \to \infty}\!\!\!\! && 
g^{(\infty)}_{ab} = \phi_{\infty}^{-1} \, \tilde{g}^{(\infty)}_{ab} \, .
\eeqn
Thus the Einstein frame vacuum field equations for BD gravity, in the $|\omega| \to \infty$ limit, read
\beqn
&& \tilde{R}_{ab}-\frac{1}{2}\tilde{g}_{ab}\tilde{R} =
A^{(\infty)}_{ab} - 8\pi \, U(\tilde{\phi}_\infty) \, \tilde{g} ^{(\infty)} _{ab} \, , \\
&&
\tilde{\Box} \tilde{\phi} = \frac{dU}{d\tilde{\phi}} \Bigg|_{\tilde{\phi}_\infty} \, ,
\eeqn
with $\tilde{\phi}_\infty = \sqrt{|\omega|/8\pi} \, \log (\phi_{\infty} / \phi_0)$. Hence, if $U(\tilde{\phi})$ is a sufficiently regular potential and $\phi$ has the asymptotic behavior discussed above, then the field equations of {\em vacuum} BD gravity in the Einstein frame correspond to those of Einstein's gravity with cosmological constant $U(\tilde{\phi}_\infty)$ and a minimally coupled scalar field in the $|\omega| \to \infty$ limit. However, the GR limit of {\em vacuum} BD gravity should be {\em vacuum} GR as we turn off the additional scalar degree of freedom, thus suggesting that the GR and $|\omega| \to \infty$ limits are not equivalent in general.

This argument is rather simple, although it lacks a degree of mathematical rigor. This issue can be resolved within the framework of the thermodynamics of scalar-tensor gravity. Indeed, this formalism allows for a straightforward, yet precise, derivation of the above statement based on its Einstein frame formulation.

First, let us observe that differentiating the field redefinition \eqref{phiE} yields
$$
d\tilde{\phi} = \phi_{0}\,\sqrt{\frac{|2\omega+3|}{16\pi}}\,\frac{d\phi}{\phi} \, ,
$$
which, using the expansion in \eqref{A1}, implies
\be
\tilde{{\nabla}}_{a}\tilde{\phi}=\frac{\phi_{0}}{\phi_{\infty}}{\frac{1}{\sqrt{8\pi}}}\,\nabla_{a}\Phi+\mathcal{{O}}\left(\frac{1}{\sqrt{|\omega|}}\right) \, .
\ee
Then, from this result, taking advantage of the expansion \eqref{A2}, a simple calculation yields
\be 
- \tilde{{\nabla}}_{a} \, \tilde{\phi} \tilde{{\nabla}}^{a}\tilde{\phi} = -\left(\frac{\phi_{0}}{\phi_{\infty}}\right)^{2}\frac{1}{8\pi}\,\nabla_{a}\Phi\nabla^{a}\Phi + \mathcal{{O}}\left(\frac{1}{\sqrt{|\omega|}}\right) \, .
\ee
Therefore, recalling the definition \eqref{eq:defChem} of the Einstein-frame chemical potential we find that
\be 
\tilde{\mu} = \frac{1}{\sqrt{8\pi}}\frac{\phi_{0}}{\phi_{\infty}}\sqrt{-\nabla_{a}\Phi\nabla^{a}\Phi} + \mathcal{{O}}\left(\frac{1}{\sqrt{|\omega|}}\right) \, ,
\ee
as $|\omega| \to \infty$. Hence,
\be 
\lim _{|\omega| \to  \infty} \tilde{\mu} = \frac{1}{\sqrt{8\pi}}\frac{\phi_{0}}{\phi_{\infty}}\sqrt{-\nabla_{a}\Phi\nabla^{a}\Phi} \neq 0 \, ,
\ee
which concludes the proof.
\section{Implications for the Jordan-frame thermodynamics}
\label{sec:JFT}
The result in {\bf Theorem \ref{Thm-1}} makes the occurrence of the anomalous $|\omega| \to \infty$ limit evident in light of the Einstein frame formulation of the thermodynamics of scalar-tensor theories. In this section we discuss the reason why this result is not immediately apparent in terms of the thermodynamic analogy performed in the Jordan frame. Since the objective is not to prove a technical statement, but rather provide a physical explanation for carrying out the analysis in the Einstein frame, we will reduce the level of rigor in the following discussion.

Given our assumption on the asymptotic behavior in Eqs.~\eqref{A1} and \eqref{A2}, we find that
\be 
\label{A-nabla}
\nabla_a \phi \sim \frac{1}{\sqrt{|\omega|}} \, \nabla_a \Phi \, , \quad \mbox{as} \,\, |\omega| \to \infty \, .
\ee
Then working within the Jordan-frame formulation of the thermodynamics of scalar-tensor gravity we can compute the behavior of the temperature of gravity, following the definition in Eq.~\eqref{KT}, which yields
\be
\K \T=\frac{\sqrt{-\nabla_a\phi\nabla^a\phi}}{8\pi \phi} \sim 
\frac{\sqrt{-\nabla_a\Phi\nabla^a\Phi}}{8\pi \phi_\infty |\omega|^\alpha} \left(1 - \frac{\Phi}{\phi _\infty |\omega|^\alpha}  \right) \, , \quad \mbox{as } |\omega| \to \infty \, .
\ee
This immediately implies that
\be 
\lim _{|\omega| \to  \infty} \K \T = 0 \, ,
\ee
thus suggesting that the theory relaxes to GR in the $|\omega| \to  \infty$, seemingly contradicting the result in {\bf Theorem \ref{Thm-1}}. We show here that the violation of {\bf Theorem \ref{Thm-1}} is only apparent, and related to the reliance of the formalism on the imperfect fluid representation of the effective tensor $T^{(\phi)}_{ab}$.

Consider the BD Lagrangian density
$$
\mathcal{L} [\phi, \nabla\phi] = \phi R - \frac{\omega}{\phi}\nabla_c\phi\nabla^c\phi-V(\phi) \, ,
$$
given \eqref{A1},\eqref{A2}, and \eqref{A-nabla} we have that
$$
\mathcal{L} [\phi, \nabla\phi] \sim \phi_{\infty}R-\frac{1}{\phi_{\infty}}\nabla_c \Phi \nabla^c \Phi - V(\phi_\infty) \, , \quad \mbox{as } |\omega| \to \infty \, ,
$$
similarly to what was correctly pointed out in \cite{Nguyen:2024jgy}. In other words, in the limit $|\omega| \to \infty$ the Jordan frame formulation of {\em vacuum} BD gravity reduces to {\em vacuum} [or with a cosmological constant $V(\phi_\infty)$] Einstein gravity minimally coupled to the scalar field $\Phi$. The effective stress-energy tensor associated to $\phi$ then becomes a perfect fluid, which corresponds to a vanishing temperature in the thermodynamic formalism. Arguably, this is a {\em GR-like} behavior of the theory, which is consistent with the typical interpretation for the vanishing of the temperature of gravity.\footnote{Adhering to this interpretation then leads to the thermodynamic derivation of the Damour-Nordvedt attractor mechanism, see \cite{Faraoni:2025alq}.} Yet, the technical definition of GR limit of scalar-tensor gravity requires that any contribution from the additional gravitational scalar degree of freedom must disappear, i.e. $\phi = \mbox{constant}$. So, the fact that we have a {\em remnant} scalar field as $|\omega| \to\infty$, rather than {\em vacuum} GR, confirms that this limit does not coincide with the GR limit of BD gravity, thus resolving the {\em apparent} inconsistency of $\K\T = 0$ and $\tilde{\mu} \neq 0$ as $|\omega| \to \infty$.
\section{Conclusions}
The inequivalence of the $\omega\to\infty$ limit and the GR limit of BD gravity is a notorious technical issue that arises in scalar-tensor gravity. 

In this work we have investigated the $\omega\to\infty$ limit of {\em vacuum} BD gravity within the formalism of the thermodynamics of scalar-tensor gravity. The Einstein-frame formulation of this paradigm allows us to describe, under fairly general assumptions, the occurrence of this anomalous behavior as a consequence of a non-vanishing chemical potential $\tilde{\mu}$ for the theory as $|\omega| \to \infty$.

We further discuss the implications of this result for the Jordan frame formulation of {\em vacuum} BD gravity, disposing of an apparent inconsistency deriving from the vanishing of the ``temperature of gravity'' in the said limit. 

\section*{Acknowledgments}
A.G. is supported by the Italian Ministry of Universities and Research (MUR) through the grant ``BACHQ: Black Holes and The Quantum'' (grant no. J33C24003220006) and by the INFN grant FLAG. A.M. is partially supported by MUR under the PRIN2022 PNRR project n. P2022P5R22A. V.F. is supported by the Natural Sciences \& Engineering Research Council of Canada (Grant no. 2023--03234). This work has been carried out in the framework of activities of the National Group of Mathematical Physics (GNFM, INdAM).

\end{document}